\documentclass[aps,showpacs,PRB,twocolumn,superscriptaddress]{revtex4-1}
\makeatletter

\newcommand{\Rmnum}[1]{\expandafter\@slowromancap\romannumeral #1@}
\makeatother
\usepackage{amsmath}
\usepackage{graphicx}
\usepackage{subfigure}
\usepackage{color}
\usepackage{amsfonts}
\usepackage{tikz}
\usepackage{esint}
\usepackage{mathrsfs}

\newcommand{\bea}{\begin{eqnarray}}
\newcommand{\eea}{\end{eqnarray}}

\newcommand{\f}{\frac}
\newcommand{\vdimer}{{\vrule height0.2cm width0.05cm depth0pt}}
\newcommand{\hdimer}{{\hrule height0.05cm width0.2cm depth0pt}}

\newcommand{\verdimers}{\hbox{\vdimer \hskip 0.1cm \vdimer}}
\newcommand{\hordimers}{\hbox{\vbox{\hdimer \vskip 0.1cm \hdimer}}}

\begin{document}

\title{Global scheme of sweeping cluster algorithm to sample among topological sectors}
\author{Zheng Yan}
\email{zhengyan\_phys@foxmail.com}
\affiliation{Beihang Hangzhou Innovation Institute Yuhang, Hangzhou 310023, China}
\affiliation{Department of Physics and HKU-UCAS Joint Institute of Theoretical and Computational Physics,The University of Hong Kong, Pokfulam Road, Hong Kong SAR, China}
\affiliation{State Key Laboratory of Surface Physics and Department of Physics, Fudan University, Shanghai 200438, China}

\begin{abstract}
Local constraint is closely related to the gauge field, so constrained models are usually effective low energy descriptions and important in condensed matter physics. On the other hand, local restriction hinders the application of numerical algorithms. In addition to the computational difficulties of the constraints, the various topological sectors which cannot be connected through local operators are also one of the key computational difficulties. Taking quantum dimer model as an example in this paper, we construct a global scheme based on sweeping cluster Monte Carlo method, which can sample among different topological sectors. In principle, this method can be generalized to other models.
\end{abstract}
\maketitle
\section{introduction}
A common theme in modern many-body physics is local constraint which always arises when there is a particularly large energy scale in frustrated Hamiltonian. We usually use gauge field theory to describe them in mathematics. However, the numerical calculation of the constrained models is difficult, which directly delays our research and understanding for these many-body systems. In addition to the computational difficulties of the constraints, the various topological sectors which cannot be connected through local operators are also one of the key computational difficulties. Constructing a global update scheme to overcome the difficulty of sampling in all topological sectors is a very important task in the development of computational methods.

For example, quantum dimer models (QDMs) featured by strong geometric restrictions are effective low energy descriptions of many frustrated quantum spin systems\cite{RK1988,BFG2002,Misguich2003,Poilblanc2010,Plat2015z2,Pollmann2015z2,YCWang2017QSL,YCWang2018,GYSun2018,YanZheng2019b,yan2020triangular,QDMbook,YCWang2021vestigial,yan2022triangular,ZYloop2022,ZZrk2021}. The QDM Hamiltonian on square lattice can be written as
\begin{equation}
    H=-\sum_{\rm plaq}\left(\vphantom{\sum}|\verdimers\rangle\langle\hordimers|+\rm{H.c.}\right)
       +V\sum_{\rm plaq}\left(\vphantom{\sum}|\verdimers\rangle\langle\verdimers|+
                                             |\hordimers\rangle\langle\hordimers|\right)
\label{Hamiltonian}
\end{equation}
where the summations are taken over all elementary plaquettes of the lattice. This seemingly simple Hamiltonian contains strong geometric constraint which requires every site on the lattice to be covered by one and only one dimer. The QDM Hamiltonian on triangular and other lattices are similar to this, where they all satisfy this constraint, and are composed of kinetic energy (resonance between the dimers of a plaquette) and their potential energy.

\begin{figure}[htp]
	\centering
	\includegraphics[width=1\columnwidth]{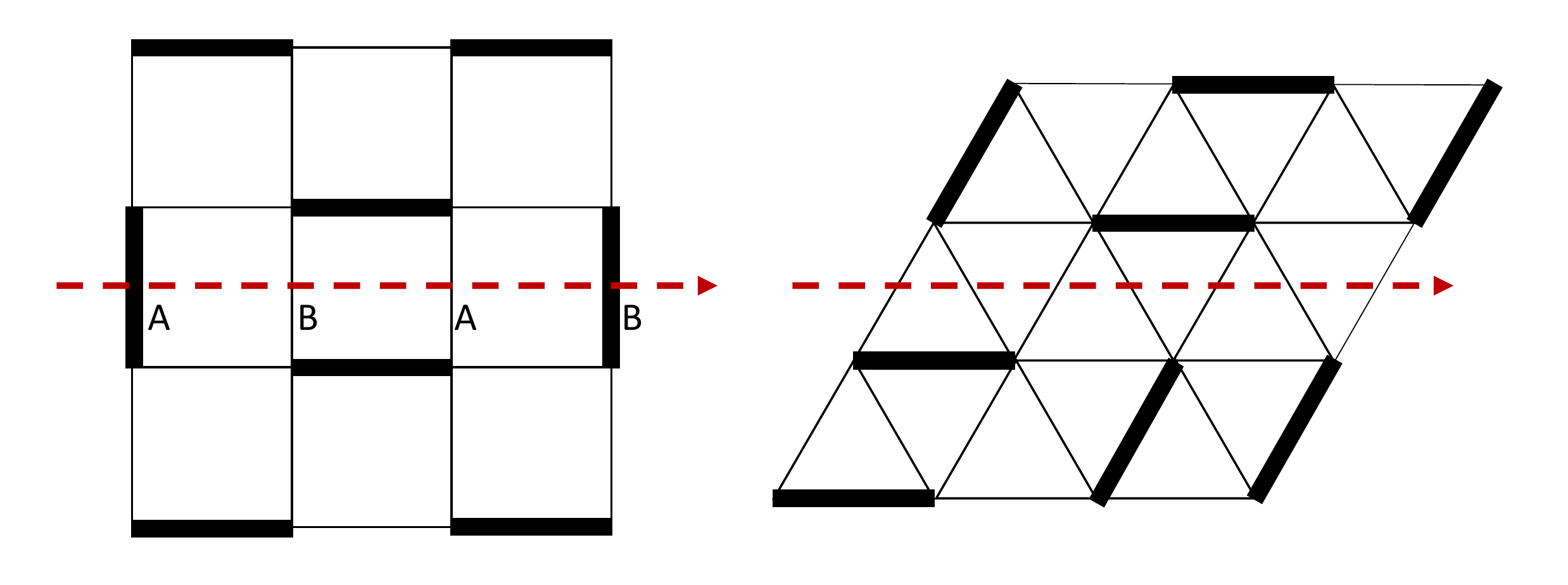}
	\caption{For the square lattice, the winding number of the x axis is defined as $W_x=N_x(A)-N_x(B)$, where the $N_x(A)$ and $N_x(B)$ are the number of dimers that the dashed line cuts on $A$ or $B$ links. The same as does the the winding number of the y axis. For the triangular lattice, winding number is defined as 0 (even)/1 (odd), when the number of dimers that are cut by the dashed line is even/odd.}
	\label{Fig1}
\end{figure}

\begin{figure}[htp]
	\centering
	\includegraphics[width=1\columnwidth]{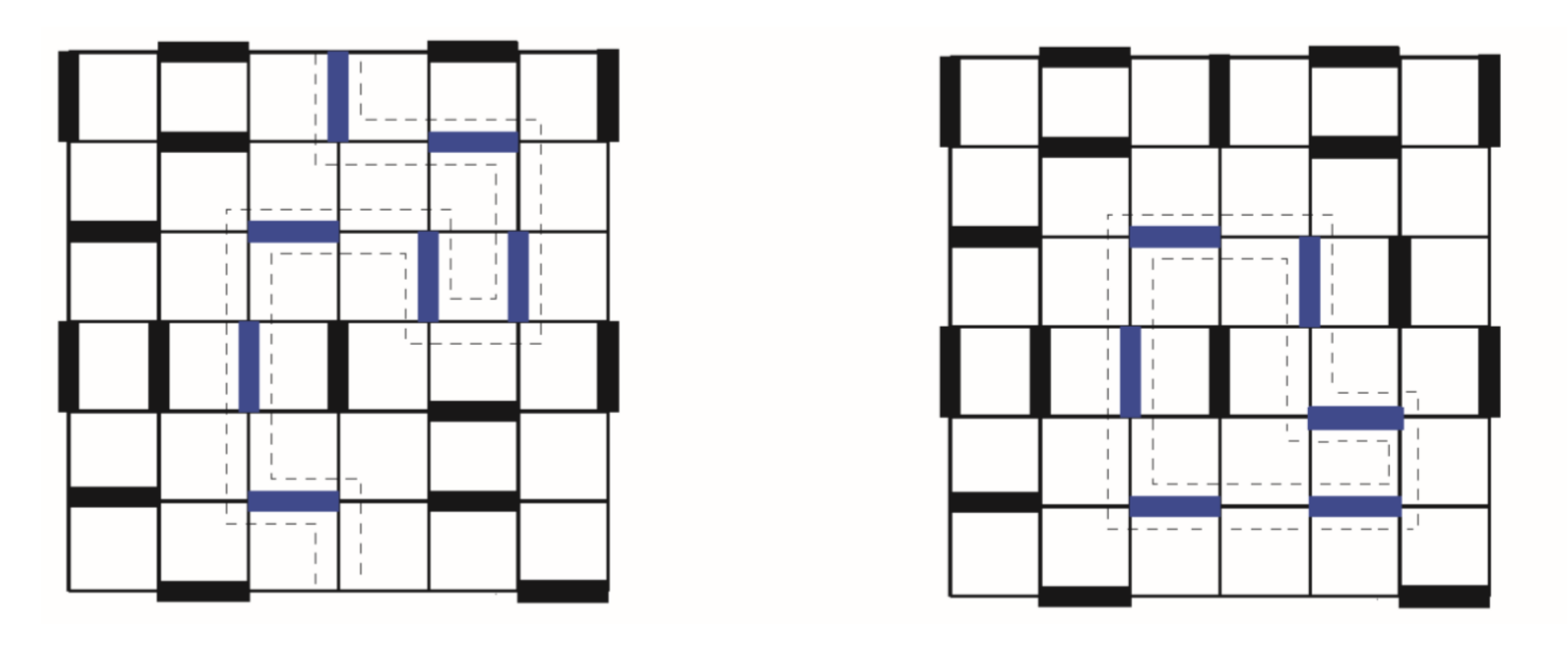}
	\caption{Left: A global loop update crosses through the boundary to change the winding number. Right: A local loop can not change the wingding number.}
	\label{Fig2}
\end{figure}

Usually, there is a $U(1)$/$Z_2$ gauge field on bipartite/nonbipartite lattice QDM due to the restrictions\cite{QDMbook}. Taking square/triangular lattices as examples, we can define the winding number and winding (topological) sector for them as Fig.\ref{Fig1} ~\cite{QDMbook,XF1,XF2,XF3}. Different winding numbers describe different winding sectors on bipartite lattice. And the parity of the winding numbers determines the different winding sectors on nonbipartite lattice. It is obvious that local operators such as the Hamiltonian operator can not change the winding sector, while only a global loop which crosses through the boundary can, as shown in Fig.\ref{Fig2}.

We developed an efficient and exact quantum Monte Carlo (QMC) based on the stochastic series expansion (SSE) method~\cite{Sandvik1991,Sandvik1999,loop2,sandvik2019stochastic,Nisheeta2021}, called ''sweeping cluster'' algorithm (SCA)~\cite{ZY2019}, which automatically satisfies the local constraint. Before sweeping cluster method was applied in world-line Monte Carlo (MC) algorithm, there is only projector MC which obeys the constraints and could be used for calculation on QDMs~\cite{OFS2005walk,OFS2005,OFS2006}. However, the projector MC for QDMs has some drawbacks, e.g., it is not effective when the parameter interval is away from Rokhsar-Kivelson (RK) point. Moreover, the projector method still lacks a cluster update to improve its efficiency. Comparing with projector MC, SCA solved the cluster update problem for constrained systems. However, it still only works in one winding sector, the same as projector MC, which needs to be improved. In many frustrated magnet cases, it is important to change the topological sectors, such as the phase diagram study of triangular lattices QDM~\cite{Ralko2005TR}, Cantor deconfinement~\cite{fradkin2004bipartite,Mosseri2015eQDM}, the finite temperature study of QDM~\cite{nogueira2009renormalization}, and quantum annealing~\cite{yan2021sweeping}, fragmental systems~\cite{XFZ2013PRL,ZYhqdm2022,CKZfracton2022}.

In this paper, we further develop a global scheme based on sweeping cluster algorithm to enable sampling between different winding sectors. In principle, this method works on any lattice QDM and can be generalized to other constrained models. In the following, we use the QDM on square lattice as examples to elaborate the details of this algorithm and provide simulations as benchmarks.

\section{a simple introduction of sweeping cluster update}
\begin{figure}[htb]
\includegraphics[width=1\columnwidth]{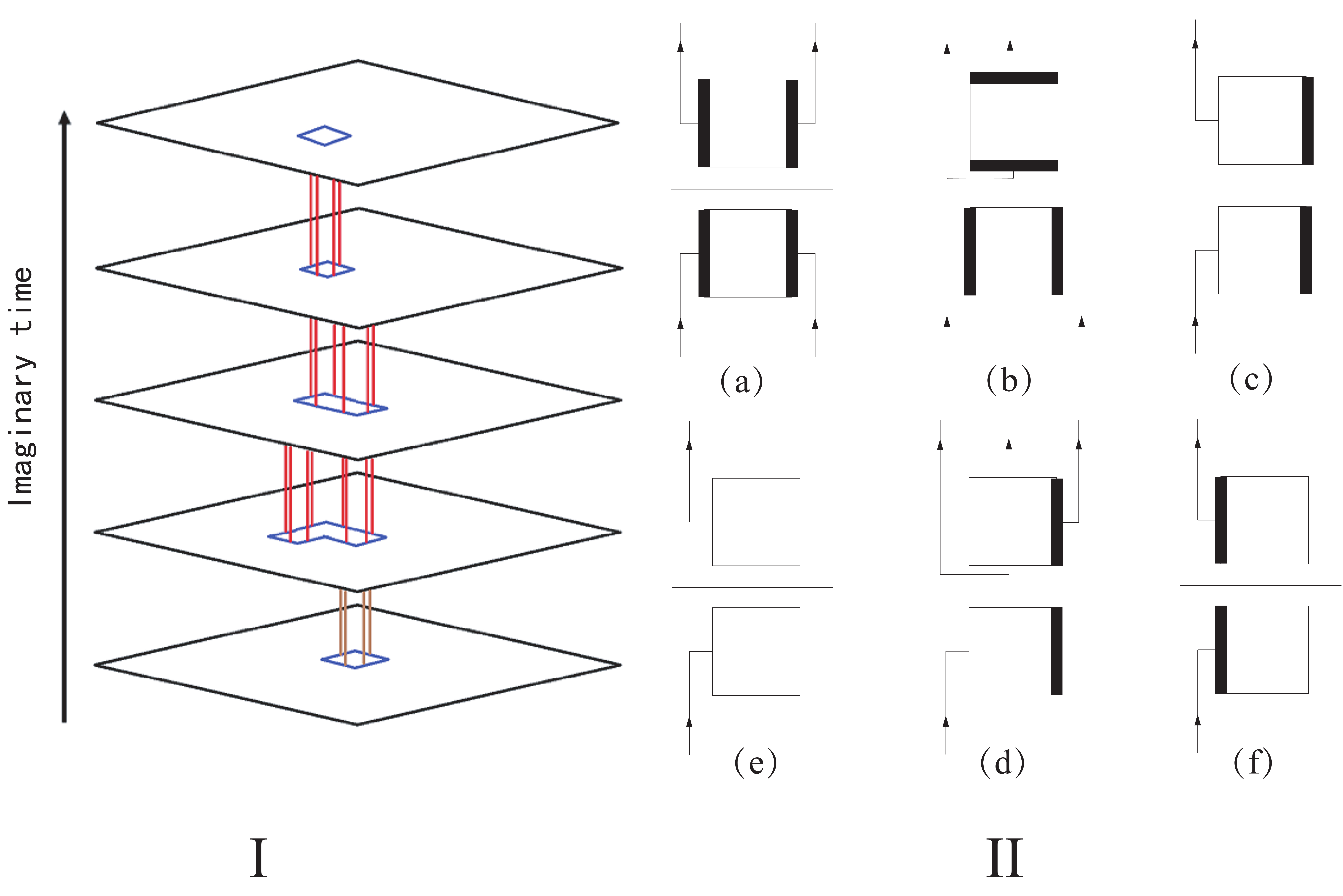}
\caption{(I) Schematic diagram of an update for quantum dimer models. Each imaginary time surface is a classical dimer configuration. Red lines are update-lines of world-line QMC. The blue loops are the intersection of all imaginary time update lines and each imaginary time surface which are the same as the loop in right part of Fig.2. (II) Some examples of the vertices and their update prescriptions. The horizontal bar represents the full plaquette operator $H_p$ and the lines of the squares represent the dimer states (thick and thin lines for dimer 1 or 0) on either side of the operator. Update-lines are shown as lines with an arrow. (c) and (d) are different updates of the same configuration. This figure is from Ref.~\cite{ZY2019}.}
\label{Fig3}
\end{figure}

We will start with a brief review of SCA based on SSE framework~\cite{ZY2019}. The key idea of SCA is sweeping the configurations one layer by one layer along imaginary time and connecting two close configurations with update-lines and an operator in the rule of SSE, to keep the constraint as Fig.\ref{Fig3}(I).

We write the Hamiltonian in terms of plaquette operators $H_p$, $H=-\sum_{p=1}^{N_p}H_p$, where $p$ labels a specific plaquette on the lattice. The plaquette operators are further decomposed into two operators: $H_p = H_{1,p} + H_{2,p}$, where $H_{1,p}$ is diagonal and $H_{2,p}$ is off-diagonal:
\begin{eqnarray}
H_{1,p} & = &
-V  \left( \vphantom{\sum} | \verdimers \rangle \langle \verdimers |+| \hordimers \rangle \langle \hordimers |\right) + V + C,
\label{hb1} \\
H_{2,p} & = & \left(  \vphantom{\sum} | \verdimers \rangle \langle \hordimers | + | \hordimers \rangle \langle \verdimers | \right).
\label{hb2}
\end{eqnarray}
Here we have subtracted a constant $N_p(V+C)$ from Eq.~(\ref{Hamiltonian}). The constant $V+C$ should make all matrix elements of $H_{1,p}$ positive which means $C> {\rm max}(-V,0)$. We will choose $C=1$ in this article for convenience.

The powers of $H$ in the SSE of the partition function $Z$ can be expressed as sum of products of the plaquette operators (\ref{hb1}) and (\ref{hb2}). Such a product is conveniently referred to by an operator-index sequence: $S_n = [a_1,p_1],[a_2,p_2],\ldots,[a_n,p_n]$, where $a_i \in \lbrace 1,2\rbrace$ corresponds to the type of operator ($1$=diagonal, $2$=off-diagonal) and $p_i \in \lbrace
1,\ldots,N_p\rbrace$ is the plaquette index. It is also convenient to work with a fixed-length operator-index list with $M$ entries and to include the identity operator $[0,0]$ as one of the operator types.

The expanded partition function takes then the same form as the SSE in the spin models~\cite{Sandvik1991,Sandvik1999},
\begin{equation}
Z = \sum\limits_\alpha \sum_{S_M} {\beta^n(M-n)! \over M!}
    \left \langle \alpha  \left | \prod_{i=1}^M H_{a_i,p_i}
    \right | \alpha \right \rangle ,
\label{zm}
\end{equation}
where $n$ is the number of operators $[a_i,p_i] \not= [0,0]$.
Inserting complete sets of states between all the plaquette operators, the product can be written as a product of the following
non-zero plaquette matrix elements
\begin{eqnarray}
\nonumber
&\langle\verdimers | H_{1,p} | \verdimers \rangle  =
\langle\hordimers | H_{1,p} | \hordimers \rangle  = 1,\\
&\langle\verdimers | H_{2,p} | \hordimers \rangle  =
\langle\hordimers | H_{2,p} | \verdimers \rangle = 1,\\
\nonumber
&\langle {\rm others}   | H_{1,p} | {\rm others}   \rangle  =  1+V,
\label{matrelem}
\end{eqnarray}
where the $| {\rm others}\rangle$ here means that plaquette $p$ has 1 or 0 dimer.
Such matrix elements are depicted in Fig.~\ref{Fig3}(II) where the plaquette below(above) is the ket(bra), we also call them as vertexes in the following.
\begin{figure*}[htb!]
\includegraphics[width=\textwidth]{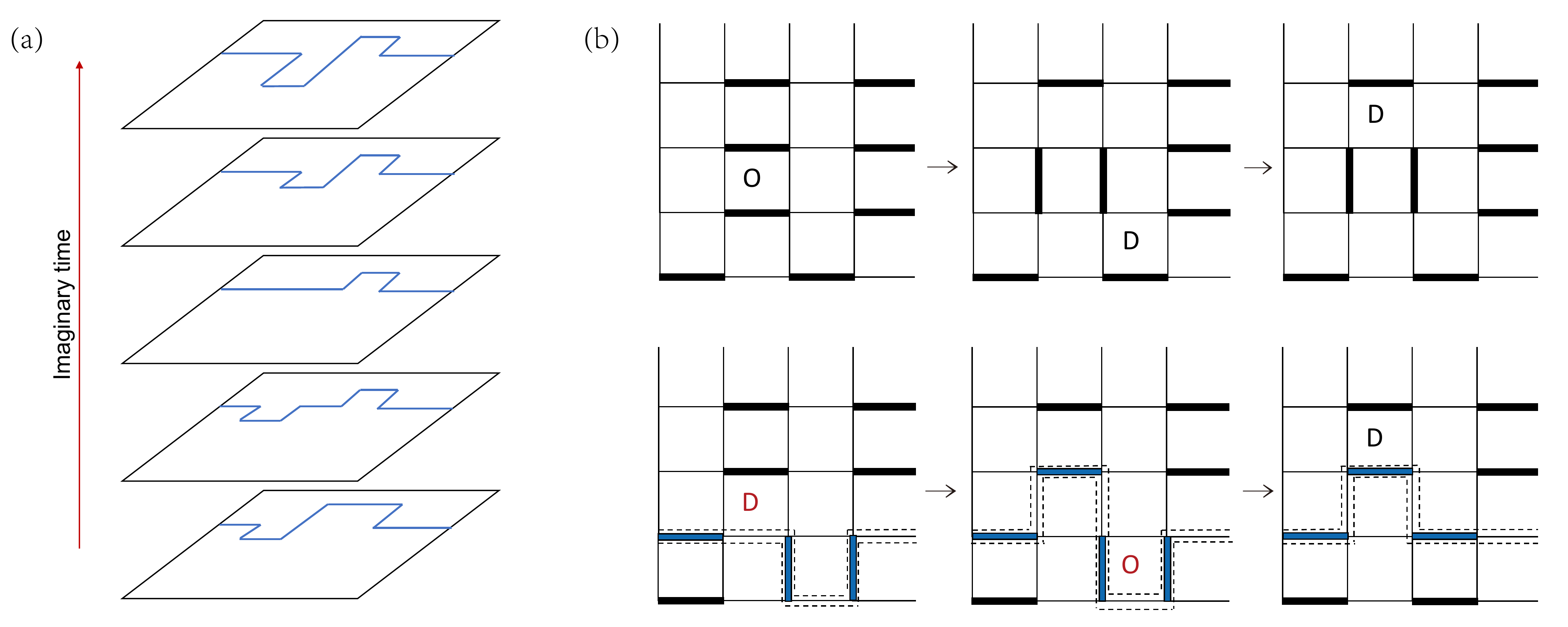}
\caption{(a) Schematic diagram of a global scheme for quantum dimer models. Each imaginary time surface is a classical dimer configuration. Here we ignore drawing update-lines of world-line QMC for convenience. The blue loops are the intersection of all imaginary time update lines and each imaginary time surface, the same as the loop in left part of Fig.2. The evolution of loop via operators can be seen clearly. (b) Configurations of QDM in imaginary time space. Each picture is a dimer configuration at a certain imaginary time, and the arrows indicate the increasing imaginary time. The D and O means a diagonal and off-diagonal operator which acts on the configuration. 1st row and 2nd row stand for dimer configurations snapshots before and after the global sweeping cluster algorithm (GSCA) update respectively. The red D or
O means the operators after update. The dashed line here means an update-line exist on the link,i.e., the dimer has to be toggled on/off. The updated dimers are blue.
}
\label{Fig4}
\end{figure*}

Diagonal update is also similar as in spin models: We accept the insertion/deletion according to the Metropolis acceptance probabilities,
\begin{eqnarray}
P_{\rm ins} & = &
{N_p\beta \langle \alpha| H_{1,p} | \alpha \rangle \over M-n },
\label{diap1} \\
P_{\rm del} & = &
{M-n+1 \over N_p\beta \langle \alpha| H_{1,p} | \alpha\rangle }.
\label{diap2}
\end{eqnarray}
The presence of $N_p$ in these probabilities reflects the fact that there are $N_p$ random choices for the plaquette $p$ in converting
$[0,0]\to [1,p]$, but only one way to replace $[1,p]\to [0,0]$ when $p$ is given. These diagonal updates are attempted consecutively for all
$1,\ldots,M$, and at the same time the state
$|\alpha \rangle$ is updated when plaquette flipping operators $[2,p]$ are encountered.

After diagonal update, we start to do cluster update. Sweeping cluster method works as follows:

Firstly, choose a flippable plaquettes(FPs) randomly no matter whether it is diagonal or off-diagonal as starting operator vertex. FP means the plaquette has two parallel dimers. Secondly, create four update-lines from every link of the plaquette, and all the lines go along one imaginary-time direction until they touch next vertex. The update-lines grown up in the imaginary-time direction will change the vertex configuration: The links touched by update-lines will create/cancel dimers as sweeping along imaginary-time.Thus the four initial update-lines rotate the two dimers of the original FP as they go along. The update-lines are extended until one or more of the update-lines hit another operator vertex from below.%

Then, after updating the plaquette beneath on the new operator vertex according to the update-lines, we need to decide how to create or destroy update-lines to update the plaquette above and continue sweeping.

For this, there are three different processes to consider: (1) The new plaquette beneath is an FP, and the old plaquette above is not an FP. We can then change the plaquette above into an FP in two ways: either the resulting vertex will either become diagonal or off-diagonal. We choose between these two possibilities shown in (c) and (d) in Fig.~\ref{Fig3}(II) with probability $1/2$. (2) The new plaquette beneath is not an FP. Then change the upper plaquette to be same as the one underneath, as shown in  (a), (b), (e) and (f) in Fig.~\ref{Fig3}(II). (3) Both the new plaquette beneath and the old plaquette above are FPs. Then there are two choices: the cluster-update ends if the number of total lines is four. If not, the four update-lines continue through the vertex and sweep on.

At the end of the sweeping cluster update, when the last four update-lines are deleted, we get a new configuration B with weight $ W_{B} $ to replace the old configuration A with weight $ W_{A} $.
To ensure detailed balance, we must invoke a Metropolis accept/reject step~\cite{Metropolis} on the whole cluster update with an acceptance probability. If we denote the number of operator vertices in configuration A with FPs on both sides by $N_{\rm FP}$, and the same amount in configuration B by $N_{\rm FP}+\Delta$, then
\begin{equation}
P_{accept}(A\rightarrow B)  =  \min( \f{N_{\rm FP}}{N_{\rm FP}+\Delta} \left( \f{2}{1+V} \right)^\Delta, 1 ).
\label{select-accept2}
\end{equation}

That is all about the original sweeping cluster method. Although it works better than the previous projector QMC methods, it can not change the winding sector while sampling.  The sampling Hilbert space of SCA is in the winding sector of the initial state forever, the same as projector MC. It means the update of SCA is local. A global scheme still needs to be developed.

\section{construct global scheme of sweeping cluster algorithm}
Comparing with projector MC used previously, SCA solved the cluster update problem for constrained systems. However, starting from a FP means all updates are derived from Hamilton's dynamics, it must be local. In fact, a configuration of QMC can be understood as imaginary time evolution of classical dimer configurations, and SCA fully samples these evolution configurations in the given sector.

In order to sample all sectors, based on the SCA, we further generalize the directed loop algorithm of N-dimensional classic dimer model\cite{Alet2003PRE,Alet2005PRE,Alet2005b,Alet2006PRE} to N+1-dimensional QDM. We keep the same diagonal update and cluster update as in original SCA method. After that, we will add a new step into the global method which can change winding sectors.
\begin{figure}[htb]
\includegraphics[width=0.9\columnwidth]{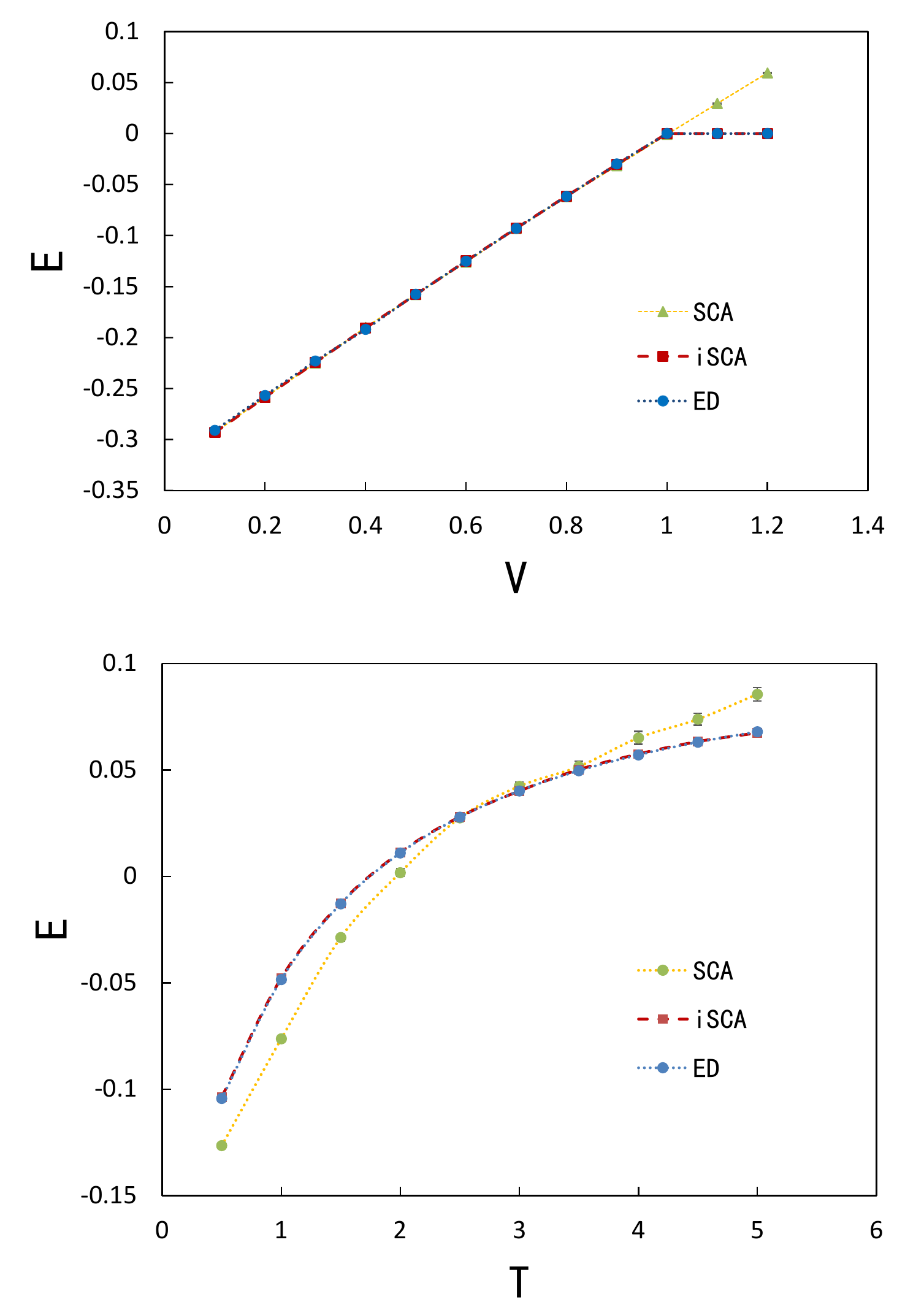}
\caption{Compare energy per plaquette of QDM by three methods on $4\times4$ square lattice: Excat diagonalization (ED) method which works in all topological sectors, original SCA QMC which only works in a certain sector (we choose (0,0) sector here, i.e., columnar sector) and GSCA which samples in all sectors. (a) At $T=0.01$, we see the ground state of ED and GSCA transfers from columnar sector to staggered sector in $V>1$ while SCA always stays in columnar sector. Both ED and GSCA work well near the topological first-order phase transition point (V=1). (b) At $V=0.5$, the energy of GSCA which contains information of all sectors matches well with the correct result of ED, but SCA is not right at finite temperature because the algorithm is confined in columnar sector of initial state. Note: The error bars are small and many are obscured by data points.}
\label{Fig5}
\end{figure}

At very high temperature, there is an easy solution to change sectors. After cluster update, walk randomly on free links until a loop is formed as Fig.\ref{Fig2} and flip all links on it in whole imaginary time. The free links here mean that there is no operator on it along imaginary time. It is worth noting that the loop here must be connected via one dimer and one empty link staggerly, and it may be local or global as Fig.\ref{Fig2}. However, when temperature becomes not high enough, there are less free links and it becomes impossible to connect a loop. So we need to construct an effective global update method as Fig.\ref{Fig4} shown: construct a loop no matter links are free or not, and update the loop in whole imaginary time based on SCA rule. Its details are as follows.

Firstly, choose a dimer configuration at initial imaginary time and construct a directed loop as done in classical dimer model\cite{Alet2005b}. We start the directed loop at a randomly chosen site and go through links with and without dimer one by one until it comes back to the starting point and closes. The loop may be local or global in this step; nevertheless, repeat until getting a global loop if you want to improve the effect of changing winding sectors. Now we get a random loop in this dimer configuration as shown in Fig.\ref{Fig2}. Then all the links of this loop create update-lines instead of plaquette in original SCA and begin to sweep the configurations along imaginary time direction. Here we should do it under a modified rules as below.
\begin{figure*}[htb]
\includegraphics[width=0.9\textwidth]{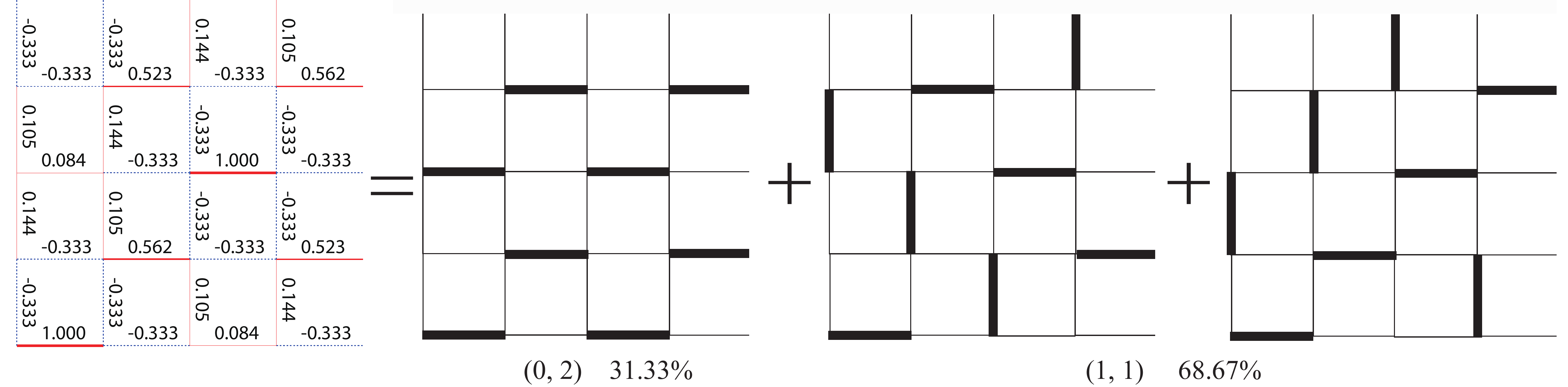}
\caption{Left one is the correlation function of QDM on $4\times4$ square lattice at $V=2,~T=0.01$. The data are rounded to three decimal places, and the errors are in/after the fourth decimal place. It can be decomposed into those staggered configurations on the right. It is worth noting that there are two kinds of staggered configurations corresponding to different winding sectors (0, 2) and (1, 1). For convenience, here we do not distinguish between the positive and negative of the winding number, and do not distinguish between (a, b) and (b, a). In the 40,000 Monte Carlo steps we simulated, they accounted for 31.33\% and 68.67\% respectively.}
\label{corr}
\end{figure*}

Then, all the update-lines should go on without stopping until meeting vertex. There are three different processes to consider at the visited vertex:(1) The new plaquette beneath is a FP, and the old plaquette above is not a FP. We can then change the plaquette above into an FP in two ways: the resulting vertex will become either diagonal or off-diagonal. We choose between these two possibilities shown in (c) and (d) in Fig.~\ref{Fig3}(II) with probability $1/2$. (2) The new plaquette beneath is not an FP. Then change the upper plaquette to be same as the one underneath, as shown in  (a), (b), (e) and (f) in Fig.~\ref{Fig3}(II). (3) Both the new plaquette beneath and the old plaquette above are FPs. The four update-lines continue through the vertex and sweep on.

Lastly, if all the update-lines close at the directed loop which we constructed at the initial floor of imaginary time, the cluster is finished. We accept it via the probability $P_{accept}(A\rightarrow B)  =  \min\left(\left( \f{2}{1+V} \right)^\Delta, 1 \right)$ which is different with Eq.(\ref{select-accept2}). If some update-lines do not close, go on sweeping until they return to the initial floor again and close. We can set a truncation number N: When they return to the initial floor for the N-th time, give up the update if update-lines are still not closed. At low temperature, N=1 is always enough and N can be set larger at higher T.
\begin{figure}[htb]
\includegraphics[width=0.8\columnwidth]{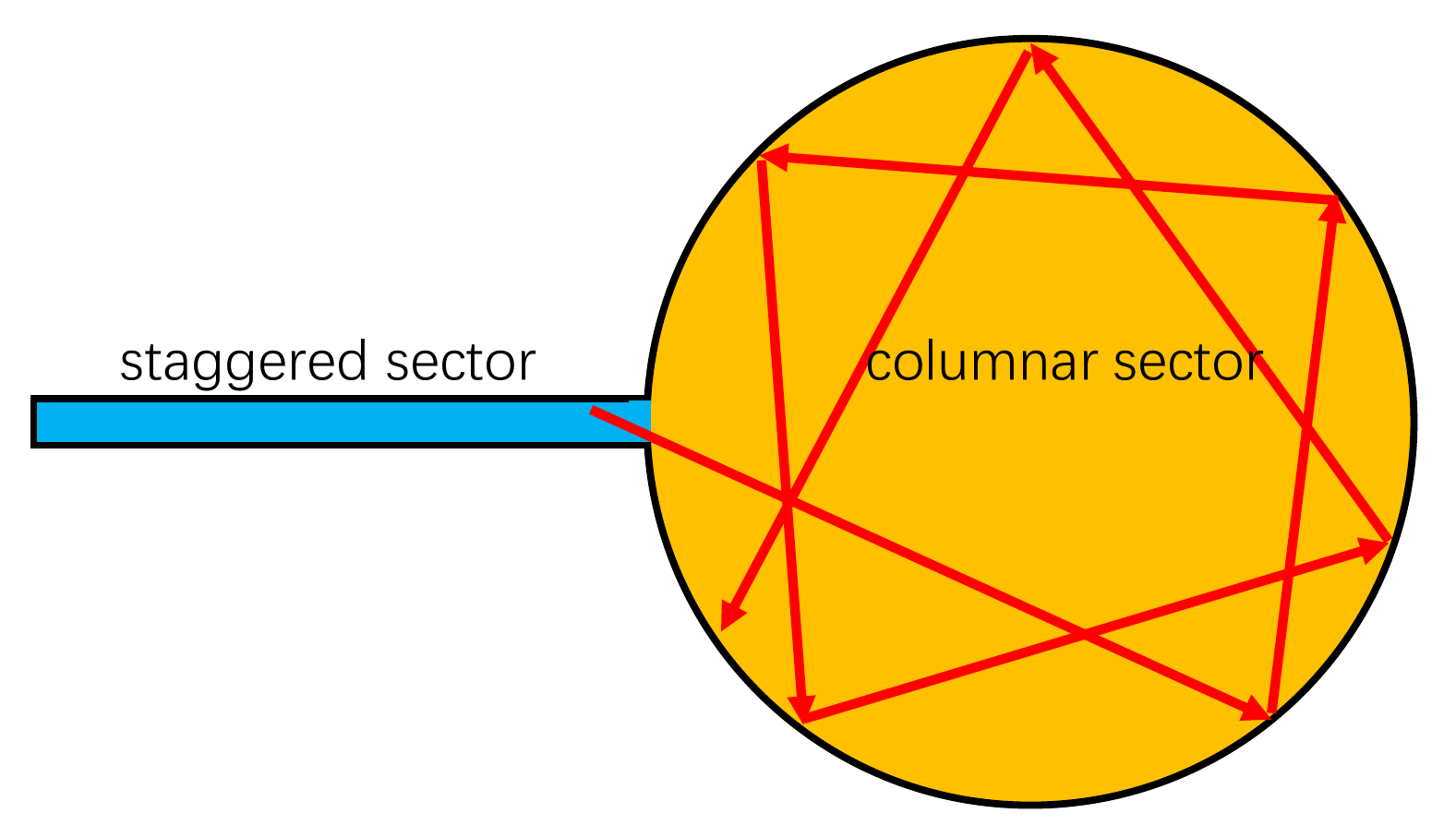}
\caption{Because the size of Hilbert space of different sectors are different, it is easy to randomly walk from staggered sector to columnar, but hard reversely, especially in large system sizes.}
\label{stag-col}
\end{figure}

Through the global scheme of sweeping cluster algorithm (GSCA) in this step, we have generalized the classical directed loop algorithm to a high-dimensional space. On the one hand, our starting update element is no longer a flippable plaquette, but can be a loop of any size (flippable plaquette is a special case, the smallest loop). On the other hand, if this loop walks across torus and is non-contractible (non-local), then we achieve a sampling which walks in different winding sectors. As the schematic diagram Fig.\ref{Fig4} (a) shown, Hamiltonian operators evolve the directed loop layer by layer along imaginary time. The GSCA updates whole configurations via Markov sampling, as an example shown in Fig.\ref{Fig4} (b).

We will provide some benchmarks as follows. In the case of at finite temperature or on different sides of $V=1$ at $T=0$, QMC should sample in different winding sectors. At low temperature, the sector of ground state when $V<1$ is $(0,0)$, i.e. columnar case, and it becomes staggered state which is in other sectors during $V>1$. It's known that the energy of staggered phase without any parallel dimers should be zero which can be a strong criterion. As Fig.\ref{Fig5} (a) shown, we see the ground state of exact diagonalization (ED) and GSCA transfers from columnar sector to staggered sector in $V>1$ while SCA always stays in columnar sector we have given through initial state. Both ED and GSCA work well near the topological first-order phase transition point (V=1). As another benchmark, we compare the QDM energy obtained by ED method, SCA and GSCA of QMC method on $4\times4$ lattice at finite temperature as Fig.\ref{Fig5} (b) shown. At finite temperature, the QMC samplings need to work in all sectors to get the correct results. As shown, the GSCA results matches well with ED. The results of both GSCA and ED are closer to zero than SCA, because staggered sector and other sectors with fewer FPs cause the energy of system to be closer to 0. On the other hand, SCA always samples in the columnar sector we have given through initial state, so it gains incorrect energy. The ED we used here is the basic full diagonalization approach.

Furthermore, we measured the correlation function as the left term of Fig.\ref{corr} shown. The correlation function is defined as $C(r)=\frac{\langle D_0D_r\rangle -\langle D_0\rangle^2}{\langle D_0\rangle -\langle D_0\rangle^2}$, $D_i=1$ (0) while there is a (no) dimer on the link. It is worth noting that there are kinds of staggered configurations corresponding to different winding sectors (0, 2) and (1, 1), so we use staggered sector instead of winding sector with certain winding numbers in this article. For convenience, here we do not distinguish between the positive and negative of the winding number, and do not distinguish between (a, b) and (b, a). According to the analysis of the value of correlation function, the left term should be accumulated by the three terms on the right as Fig.\ref{corr}. And the right two of the three staggered items actually correspond to the same winding numbers (1,1). In 40000 Monte Carlo samplings, the staggered state in (0,2) and (1,1) accounted for 31.33\% and 68.67\% respectively.

People may think that the global scheme of sweeping cluster algorithm is not efficient enough, because it requires all update-lines to be closed after whole imaginary time cycle due to the periodic boundary condition (PBC) of imaginary time. Actually, this does affect the effectiveness of the algorithm, especially when the size is larger. So a simpler way is that we can generalize it to the method of projector SSE Monte Carlo method with open boundary condition of imaginary time\cite{sandvik2005ground,sandvik2010loop}.

As Fig.\ref{Fig4} shown, a random-walk loop is constructed first in a start imaginary time layer, and the Hamiltonian operators distributed over imaginary time change the shape of the loop, resulting in a 3D update-cluster. Due to the PBC of imaginary time, it requires the loops at $\tau=0$ and $\beta$ should have same shapes to close this cluster. Obviously, it is difficult to close the update-clusters when the system size is large. However, in zero temperature case, there is an open boundary condition (OBC) along imaginary time as Eq.(\ref{zmm}) shown, i.e., the left state need not be equal to the right one. It means the shapes of loops at $\tau=0$ and $\beta$ can be different, so the update-cluster works well even if the system size is very large. The cluster can certainly be gained via the zero temperature method, but it may not be closed via the finite temperature way.

Furthermore, it is worth noting that the degrees of freedom within different sectors are different. For example, there is no flippable plaquette in the staggered sector, so it has few states, while the columnar sector is the opposite. As shown in Fig.\ref{stag-col}, it means it is easy to randomly walk from staggered sector to columnar, but hard reversely. Therefore, it is more appropriate to set the initial state of the Monte Carlo simulation to the staggered state. In the next section, we use the projector SSE method and set initial state staggered to simulate QDM in different sizes to check the effectiveness of GSCA.

\section{Global sweeping cluster algorithm in projector SSE method}
The projector Monte Carlo algorithm\cite{Sugar1983} is a common numerical method for studying ground states of quantum many-body systems. In a broad sense, Green's function Monte Carlo and diffusion Monte Carlo both belong to it. Consider a state $|\Psi\rangle$ and its expansion in terms of eigenstates $|n\rangle$, n = 0, 1, ..., of some Hamiltonian H;
\begin{equation}
    |\Psi\rangle=\sum_n a_n|n\rangle
\end{equation}
Let H be the Hamiltonian of interest. Then for sufficently large $\beta$, $e^{-\beta H}$ can be used as a projection operator onto any state of this system.
\begin{equation}\label{projector}
\begin{split}
    \lim_{\beta\rightarrow \infty}e^{-\beta H}|\Psi\rangle &=\lim_{\beta\rightarrow \infty}\sum_ne^{-\beta E_n}a_n|n\rangle\\
    &=\lim_{\beta\rightarrow \infty}e^{-\beta E_0}\sum_ne^{-\beta (E_n-E_0)}a_n|n\rangle\\
    &=\lim_{\beta\rightarrow \infty}e^{-\beta E_0}a_0|0\rangle
\end{split}
\end{equation}
Then, from this expression, one can write a normalization of the groundstate wavefunction like partition function, $Z=\langle0|0\rangle$ with two projected states (bra and ket) as,
\begin{equation}
    Z=\lim_{\beta\rightarrow \infty}(\langle \Psi_L|e^{-\beta H})e^{-\beta H}|\Psi_R\rangle=\lim_{\beta\rightarrow \infty}\langle \Psi_L|e^{-2\beta H}|\Psi_R\rangle
\end{equation}
It is worth noting that in the actual implementation, we only need to randomly give the $\Psi_L$ state and generate the entire propagator according to the rules of SSE.
\begin{figure}[htb]
\includegraphics[width=1\columnwidth]{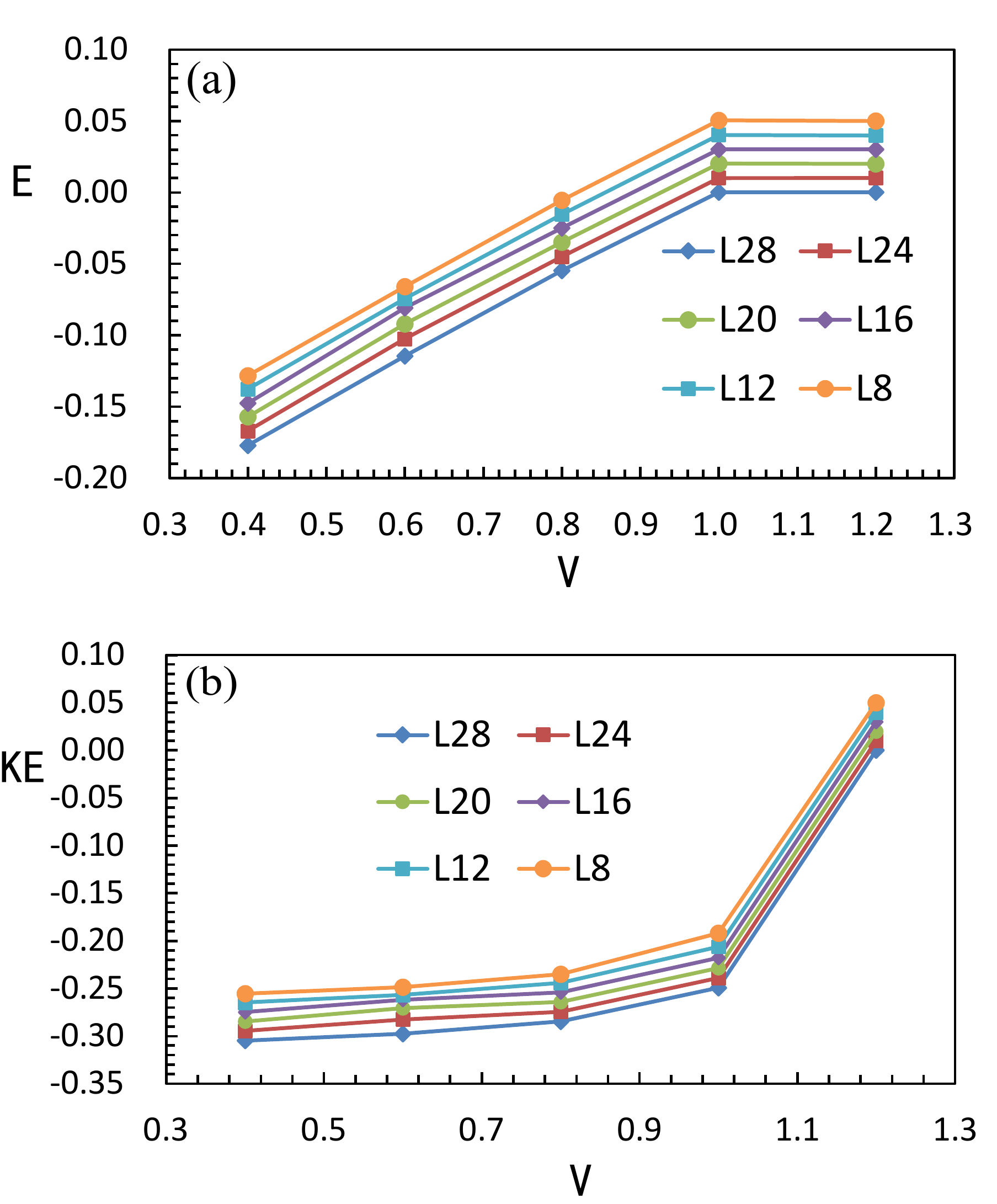}
\caption{In different sizes, the relationship of energy/kinetic energy per plaquette and parameter V. Error bars are smaller than the data points. In order to avoid overlapping of data of different sizes, we have made corresponding shifts according to sizes. The shift value is $+0.01\times(28-L)/4$. (a) shows a clear turn at $V=1$, it means the ground state jumps between columnar and staggered sectors. Energy is equal to 0 in error bar region when $V\ge 1$. (b) shows the kinetic energy is equal to 0 exactly while $V>1$, which is a strong evidence for staggered phase without any flippable plaquette. At $V=1$, kinetic energy is not 0 obviously and energy of (a) is 0, which matches with exact solution of ground state at RK point.}
\label{Fig6}
\end{figure}

For convenience, in the following we use $\beta$ instead of $2\beta$. In order to represent the normalization as a sum of weights, $Z = \sum_x W(x)$, we use Handscomb's power series expansion~\cite{Handscomb} and SSE framework~\cite{Sandvik1991,Sandvik1999} to rewrite it as,
\begin{equation}
Z = \sum\limits_{\Psi_L \Psi_R} \sum_{S_M} {\beta^n(M-n)! \over M!}
    \left \langle \Psi_L  \left | \prod_{i=1}^M H_{a_i,p_i}
    \right | \Psi_R \right \rangle ,
\label{zmm}
\end{equation}
$S_M = [a_1,p_1],[a_2,p_2],\ldots,[a_M,p_M]$, where $a_i \in \lbrace 1,2\rbrace$ corresponds to the type of operator ($1$=diagonal, $2$=off-diagonal) and $p_i \in \lbrace
1,\ldots,N_p\rbrace$ is the index of position. It is convenient to work with a fixed-length operator-index list with $M$ entries and to include the identity operator $[0,0]$ as one of the operator types. And $n$ is the number of operators $[a_i,p_i] \not= [0,0]$. In this framework, we can apply the previous sweeping cluster method in SSE~\cite{ZY2019}.

Then the steps are almost the same as those in finite temperature, the only difference is that the boundary condition of imaginary time becomes open. It means the directed loop in Fig.\ref{Fig4} at $\tau=\beta$ need not be equal to ones at $\tau=0$ and update-lines needn't match and close after $\beta$ evolution. However, the price is that it can only work at zero temperature.

We set the initial state of GSCA to be a staggered state, to simulate QDM on the square lattice. A small reminder: in the SSE code, we need to set the initial cut-off length M of imaginary time of Eq.(\ref{zmm}) to a relatively large number, here we use 2000. In the Fig.\ref{Fig6}, the QMC simulation results show that the state of staggered phase can successfully enter the sector of columnar phase. We sweep the energy/kinetic energy of several size under different parameter $V$ at zero temperature. When $V>1$, the ground state is staggered phase without any parallel dimers in a plaquette, so both the energy and kinetic energy of staggered state must be 0. This staggered configuration cannot be obtained from the ground state of $V<1$ (columnar sector) through the evolution of local operators, and vice versa. Therefore we can judge whether the transition of different topological sectors is successful sectors via the changes of energy/kinetic energy near $V=1$. Fig.\ref{Fig6} (a) shows a clear turn at $V=1$, it means the ground state jumps between columnar and staggered sectors. Energy is equal to 0 in error bar region when $V\ge 1$. It is because we add constant term into the Hamiltonian as Eq.(\ref{matrelem}) shown, so plaquettes without parallel dimers also cost energy to cause statistical fluctuation. On the other hand, (b) shows the kinetic energy is equal to 0 exactly while $V>1$, it is a strong evidence for staggered phase without any flippable plaquette. At RK point $V=1$, kinetic energy is not 0 obviously but energy is 0, it matches with exact solution of ground state at RK point. We also checked the winding number during simulation, and all the samplings of region $V<1$ become columnar sector from staggered phase after thermalization. In summary, GSCA works well in the open boundary condition of imaginary time.

\section{Extension and generalization}
We will discuss why this scheme can be applied generally. In quantum Monte Carlo algorithm, sampling depends on the imaginary time evolution of Hamiltonian operators. It is impossible to connect two different topological sectors via Hamiltonian operators. Actually, it's not only for topological sectors, but also for other sectors which can not be connected to each other through Hamiltonian. If the quantum fluctuations are not ergodic to all the samples, we need to manually introduce an update scheme as in classical systems.

In fact, we can always introduce a classical update in a certain imaginary time layer to connect different sectors, because every layer of imaginary time is a classical configuration. The point is how to extend this classical update into all imaginary time layers, which can be solved by SCA. Because SCA updates the configurations along imaginary time direction, it actually evolutes the classical update-loop via Hamiltonian operators until the end of imaginary time, as Fig.\ref{Fig4} (a) shown. Thus, the classical update is extended to quantum case in this way.

All in all, this scheme gives an available way to extend classical update scheme into quantum version, and overcome the sector-localizations due to non-connection of Hamiltonian.
\begin{figure}[htb]
\includegraphics[width=\columnwidth]{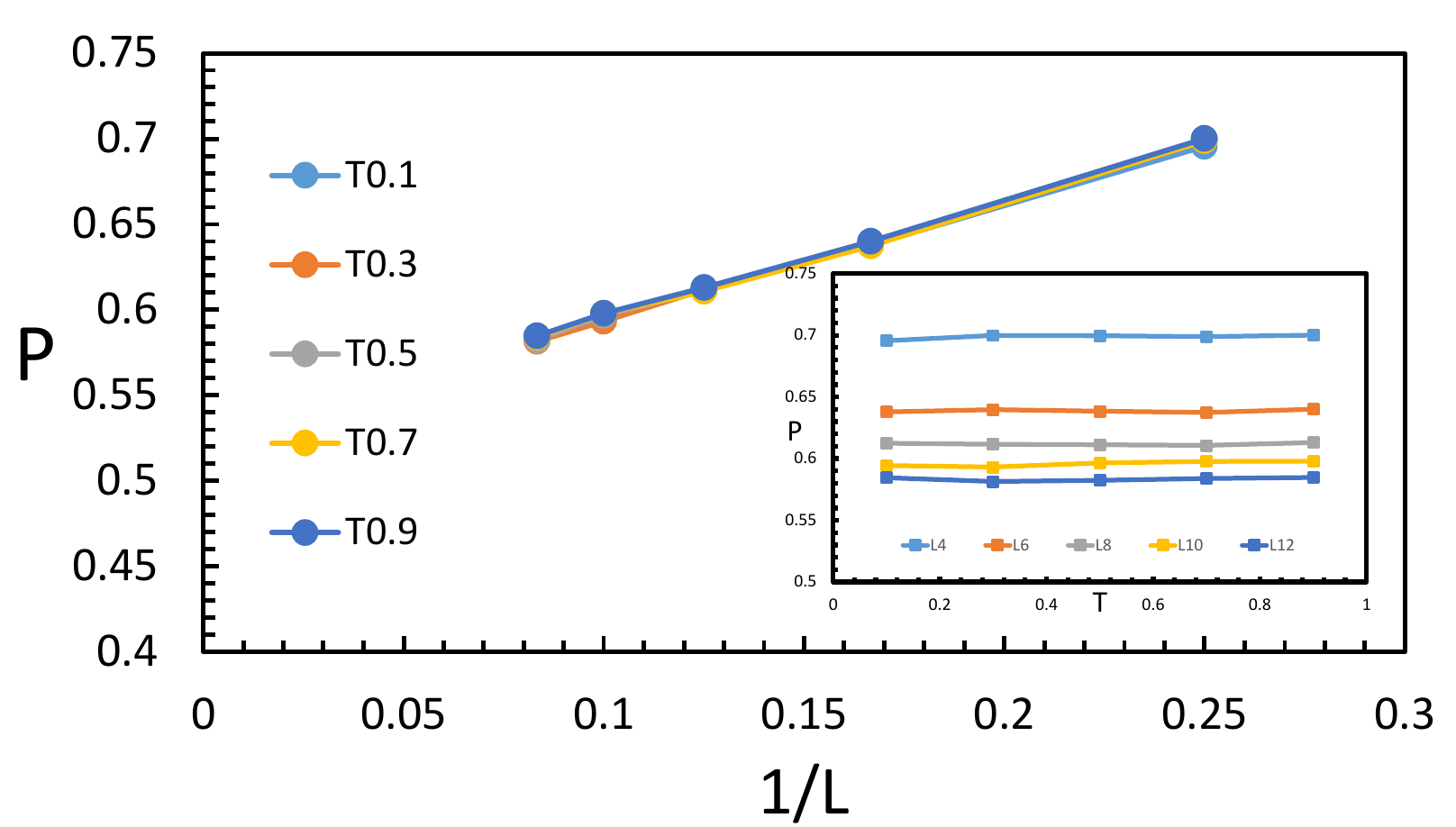}
\caption{The probability that all the update-lines of the global loop close after one round along imaginary time. We simulate different sizes and temperature cases at $V=0.5$. The inset uses the same data but changes the x-axis as $T$ instead of $1/L$.}
\label{closeprob}
\end{figure}

\section{Discussions}
Let us come back to discuss some questions about the effectiveness and range of application. In the finite temperature case, the update-lines may not close even though they go several rounds along the imaginary time. At $V=0.5$, the update-lines closed probability after one round along imaginary time of different temperature and sizes are studied in Fig.\ref{closeprob}. Although the probability falls down while the system size becomes larger, it decays very slowly and all the probabilities are greater than $55\%$. Along the linear trend, the $P$ will be larger than half even when $L \rightarrow \infty$. This suggests that the formation of a global cluster to overcome the topology is a likely option.
\begin{figure}[htb]
\includegraphics[width=\columnwidth]{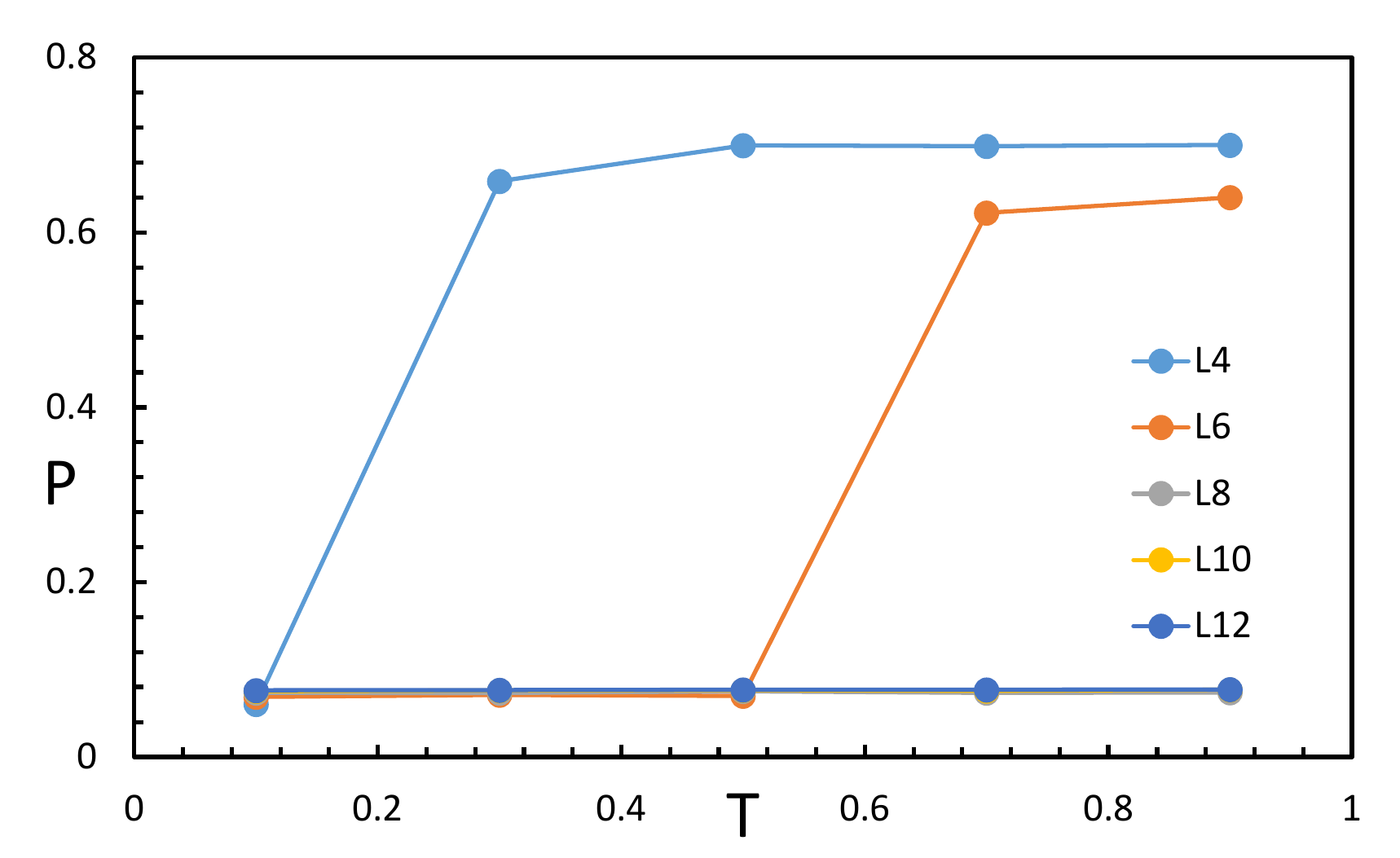}
\caption{The accepted probability of a global cluster of different temperature and sizes at V=0.5.}
\label{accptprob}
\end{figure}

Furthermore, how many global cluster can be accepted is also an important indicator. In the finite temperature case, we find that it is dependent on the system size and temperature. It can be seen that although the number for update-lines-closing in imaginary time is not small as the upper Fig.\ref{closeprob} shown, the probability for accepting these updates is not high [Fig.\ref{accptprob}], especially for large system size and low temperature. It's worth noting that if the update-lines haven't closed after one round, we regard it as not accepted. Even though the probability is not large, it may not mean the efficiency of the algorithm is bad. Because the proportion of (0,0) sector will increase with the system size, there will be more and more samplings of the sector than others. As the paper by Alet et al.~\cite{Alet2005b} mentioned, the (0,0) sector occupies nearly $100\%$ when the system size is infinite. Thus, it is a little hard to judge if the algorithm is effective in finite temperature case.
\begin{figure}[htb]
\includegraphics[width=\columnwidth]{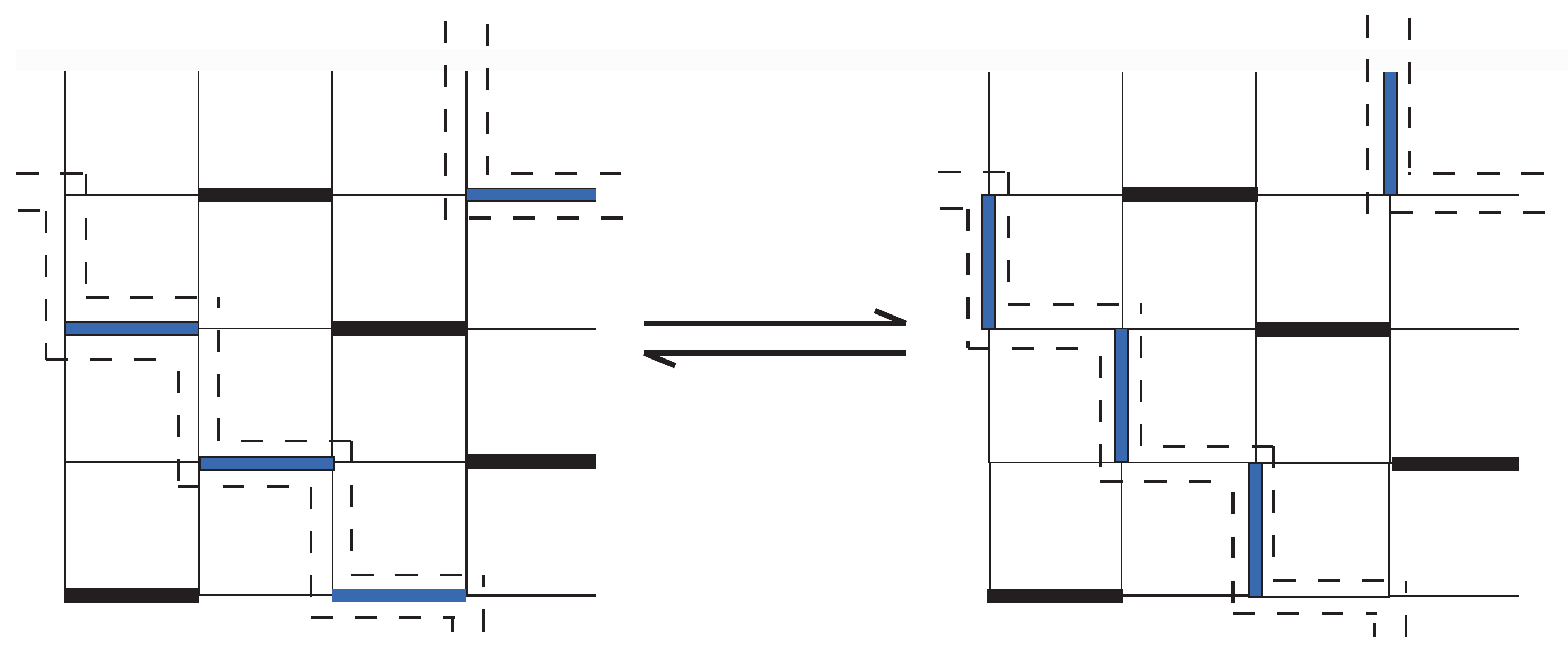}
\caption{The nearest way to connect two staggered configurations is to update the dimers along the clino-diagonal line.}
\label{stagloop}
\end{figure}

For the zero temperature case, as discussed above in the Fig.\ref{stag-col}, it is important to set initial state in a small sector, such as staggered phase. Using this trick, it is easy to find the lowest-energy sector via global updates. However, can this scheme visit all the degenerate staggered states when $V>1$? It is hard to be ergodic, especially for large size. At $V=1.5$, the probability of sector-changes is about $0.3$ when $L=4$, but it falls down to about $0.0002$	quickly when $L=10$. The reason is that it will cost energy in the intermediate processes if the state goes into other non-staggered sectors from staggered phase and then falls into another stagger, so this way will be rejected. The nearest way to connect two staggered configurations is to update the dimers along the clino-diagonal line as Fig.\ref{stagloop}. The probability of this becomes nearly zero while the system size increases hugely. A possible solution is to introduce a transcendental choice when creating the classical loop in the initial layer of imaginary time. As in classical dimer model case~\cite{Alet2003PRE,Alet2005PRE,Alet2005b,Alet2006PRE} (only V term), it favors more/less plaquettes with two dimers when the loop grows up at negative/positive $V$. In the quantum case, we can modify the judge-condition from $V=0$ to $V=1$. But it requires that we have known the staggered phase arises while $V>1$.

\section{Conclusions}
Numerical study of constrained model is important and notoriously difficult, especially as the topology forbids ergodicity. We develop a global scheme of sweeping cluster SSE method to overcome the sampling between different topological sectors. The technique keeps the geometric restrictions via sweeping vertices in imaginary-time order and achieves sampling in all winding sectors. In principle, this method works on any lattice QDM and can be generalized to other constrained models.

\section{Acknowledgements}
I wish to thank Olav Sylju{\aa}sen, Jie Lou, Yan Chen, Chenrong Liu, Ruizhen Huang, Zheng Zhou, Xue-Feng Zhang, Anders Sandvik and Zi Yang Meng for fruitful discussions. Especially thank my good friend Mao Hong very much, who helped me improve the fluency of this paper. In addition, I also want to thank my parents who took care of my wife while she was pregnant, so that I can focus on scientific research. Acknowledge my wife--Ms. LOU's gift, the birth of my son -- YAN Gejie (the name means "exact solution" in Chinese). The author acknowledges Beijng PARATERA Tech Co.,Ltd. for providing HPC resources that have contributed to the research results reported within this paper.

%

\end{document}